\documentclass[conference]{IEEEtran}
\IEEEoverridecommandlockouts
\usepackage{cite}
\usepackage{amsmath,amssymb,amsfonts}
\usepackage{algorithmic}
\usepackage{graphicx}
\usepackage{textcomp}
\usepackage{amsmath}
\usepackage{array}
\usepackage{multicol}
\usepackage{rotating}
\usepackage{adjustbox}
\usepackage{multirow}
\usepackage{makecell}
\usepackage{array}
\usepackage{longtable}
\usepackage{caption}
\usepackage{xcolor}
\usepackage{tabularx} 
\usepackage{booktabs} 
\usepackage{multirow} 
\usepackage{siunitx} 
\def\BibTeX{{\rm B\kern-.05em{\sc i\kern-.025em b}\kern-.08em
    T\kern-.1667em\lower.7ex\hbox{E}\kern-.125emX}}
    
\begin{document}

\title{Fine-tuned Transformer Models for Breast Cancer Detection and Classification \\

}

\author{\IEEEauthorblockN{1\textsuperscript{st} Showkat Osman}
\IEEEauthorblockA{\textit{Dept.of ECE} \\
\textit{North South University}\\
Dhaka, Bangladesh \\
showkat.osman@northsouth.edu}
\and
\IEEEauthorblockN{2\textsuperscript{nd} Md. Tajwar Munim Turzo}
\IEEEauthorblockA{\textit{Dept.of ECE} \\
\textit{North South University}\\
Dhaka, Bangladesh \\
munim.turzo@northsouth.edu}
\and
\IEEEauthorblockN{3\textsuperscript{rd} Maher Ali Rusho\textsuperscript{*}}
\IEEEauthorblockA{
NMR Spectroscopist\\
Lassonde School of Engineering York University \\
Toronto, Canada \\
alirusho@yorku.ca}
\and
\IEEEauthorblockN{4\textsuperscript{th} Md. Makdir Haider}
\IEEEauthorblockA{\textit{Dept.of EEE} \\
\textit{Ahsanullah University of Science and Technology}\\
Dhaka, Bangladesh \\
makid94haider@gmail.com}
\and
\IEEEauthorblockN{5\textsuperscript{th} Sazzadul Islam Sajin}
\IEEEauthorblockA{\textit{Dept.of ECE} \\
\textit{North South University}\\
Dhaka, Bangladesh \\
sazzadul.sajin@northsouth.edu}
\and
\IEEEauthorblockN{6\textsuperscript{th} Ayatullah Hasnat Behesti}
\IEEEauthorblockA{\textit{Dept.of ECE} \\
\textit{North South University}\\
Dhaka, Bangladesh \\
ayatullah.behesti@northsouth.edu}
\and
\IEEEauthorblockN{7\textsuperscript{th} Ahmed Faizul Haque Dhrubo}
\IEEEauthorblockA{\textit{Dept. of ECE} \\
\textit{North South University}\\
Dhaka, Bangladesh \\
ahmed.dhrubo@northsouth.edu}
\and
\IEEEauthorblockN{8\textsuperscript{th} Md. Khurshid Jahan}
\IEEEauthorblockA{\textit{Dept.of ECE} \\
\textit{North South University}\\
Dhaka, Bangladesh \\
khurshid.jahan@northsouth.edu}
\and
\IEEEauthorblockN{9\textsuperscript{th} Mohammad Abdul Qayum}
\IEEEauthorblockA{\textit{Dept.of ECE} \\
\textit{North South University}\\
Dhaka, Bangladesh \\
mohammad.qayum@northsouth.edu}
}

\maketitle

\begin{abstract}
Breast cancer is still the second top cause of cancer deaths worldwide and this emphasizes the importance of necessary steps for early detection. Traditional diagnostic methods, such as mammography, ultrasound, and thermography, which have limitations when it comes to catching subtle patterns and reducing false positives. New technologies like artificial intelligence (AI) and deep learning have brought about the revolution in medical imaging analysis. Nevertheless, typical architectures such as Convolutional Neural Networks (CNNs) often have problems with modeling long-range dependencies. It explores the application of visual transformer models (here: Swin Tiny, DeiT, BEiT, ViT, and YOLOv8) for breast cancer detection through a collection of mammographic image sets. The ViT model reached the highest accuracy of 99.32\% which showed its superiority in detecting global patterns as well as subtle image features. Data augmenting approaches, such as resizing croppings, flippings, and normalization, were further applied to the model for achieving higher performance. Although there were interesting results, the issues of dataset diversity and model optimization which present new avenues of research are also still present. Through this study, the crystal potential of transformer-based AI models in changing the detecting process of breast cancer and, thus, to patients' health, is suggested.
\end{abstract}


\section{Introduction}
Bangladesh, with a population of approximately 173.97 million as of 2024 (Worldometer), faces significant health challenges, with nearly half of the population comprising women\cite{b1}. Each year, a large number of people suffer from various diseases, with women disproportionately affected due to societal norms that inhibit them from openly discussing health issues. In many cases, women refrain from sharing their medical concerns, even with family members or healthcare providers, fearing stigmatization and being perceived as a burden. As a result, numerous women succumb to preventable conditions due to delayed diagnosis and treatment. Breast cancer, in particular, stands as one of the most critical health issues faced by women, both globally and in Bangladesh.

Breast cancer is the second leading cause of cancer-related death worldwide. It occurs when breast cells undergo mutations, becoming cancerous and multiplying to form tumours. Although the disease predominantly affects women, especially those assigned female at birth (AFAB) and aged 50 and older, it can also impact men and younger individuals\cite{b2}. Approximately 80\% of breast cancer cases are invasive, meaning the tumour may spread beyond the breast to other parts of the body\cite{b3}. Breast cancer is classified into common and less common types. Common types include invasive ductal carcinoma, lobular breast cancer, and ductal carcinoma in situ. Less common types include triple-negative breast cancer, inflammatory breast cancer, and Paget’s disease of the breast\cite{b3}.

Recognizing the early signs and symptoms of breast cancer is crucial for effective treatment. Common symptoms include changes in breast size, shape, or contour, persistent lumps or thickening in the breast or underarm, and discharge from the nipple, which may be blood-stained or clear. Early detection significantly improves treatment outcomes. Various imaging techniques are available for screening and diagnosing breast cancer, with mammography, ultrasound, and thermography being the most prevalent. Mammography, which uses X-rays, is the gold standard for early detection, while magnetic resonance imaging (MRI) is employed for high-risk individuals. Physical examinations, such as clinical breast exams performed by healthcare professionals and self-examinations, also play a vital role in early detection. Women’s awareness of changes in their breasts, such as alterations in size, lumps, or pain, is essential in prompting timely medical consultation.

\section{Literature Review}
Breast cancer detection has been studied widely; researchers have employed different imaging modalities along with machine learning techniques in enhancing diagnostic accuracy. This section encompasses an appraisal of all significant contributions in the field by focusing on central methodologies and all comparative strengths. Sadoughi et al.\cite{b4} used breast cancer detection methods that incorporated mammography, ultrasound, and thermography as imaging techniques. They've reported a high level of accuracy while using the SVM models and analyzing more than 40 datasets. Texture, edge detection, and fuzzy methods were emphasized in their approach to highlight the strengths and weaknesses of each modality.

Prannoy and Saravanakumar\cite{b5} developed a CAD system to class microcalcifications in mammograms as benign, normal, or malignant. This method employed Principal Component Analysis (PCA) for dimensionality reduction and included approximately 3,000 extracted cases from the DDSM database, indicating great promise in early tumor detections. Neural networks (NNs) are again another research line to be explored in breast cancer detection,Mekdy et al\cite{b6}. Neural Network implemented by ANNs, BPNNs, and hybrid models aimed at reducing false positives and negatives across many imaging modalities, including mammography, ultrasound, thermal imaging, and MRI. The versatility and adaptability NNs have shown are quite astonishing and applicable for diversity in imaging. R. Guzmán-Cabrera et al.\cite{b7} presented a CAD system that uses texture segmentation and morphological operators on digital mammograms. Texture segmentation and morphology operators are used along with machine learning techniques and entropy-based methods in their system for segments and suspicious area extraction, thus getting a high accuracy to deliver a push toward more automatic diagnostic tools.

Integration of machine learning algorithm and image processing technique for breast cancer classification has been accomplished by Jasti et al.\cite{b8}. They performed implementations of models such as LS-SVM, K-Nearest Neighbors (KNN), Random Forest, and Naïve Bayes, besides AlexNet and CNN for feature extraction. This multi-level approach achieved high precision, thus demonstrating the prowess of combining traditional and deep learning methods. Uswatun, Sigit, and Yuniarti\cite{b9} used to enhance breast cancer detection through ultrasound imaging, this study has made use of the Watershed Transform Algorithm, achieving an 88.65\% diagnostic accuracy and an 11.35\% average segmentation error on a dataset of 100 ultrasound images. This one is highlighting the importance of accurate segmentation in reducing the diagnostic errors. Jahangir et al.\cite{b10} used the Vision Transformer and Convolutional Neural Networks (CNNs) to classify brain tumors, and compared the performance using precision, recall, and F1-score. Jahangir et al found that both types of models performed well. Ideally, the Vision Transformer models outperformed on the training data with perfect results 100\% precision, 100\% recall, and 100\% F1-score while the CNNs performed relatively better on the test data with precision, recall, and F1-score scores of 98.233\%, 98.271\%, and 98.234\% respectively. The project was developed in five project phases, described as 6 models were trained in the third project phase with 6 well-known transfer learning based CNN models. The authors trained four variants of the Vision Transformer (L/32, L/16, B/32, and B/16) on the same data sets. The models were developed in layers with each converter (atria) not interrelated and tested on the sample data that they use for classification. Chauhan et al.\cite{b11} did bring an innovative way to detect brain tumors, called Patch-Based Vision Transformers. Utilizing the Figshare brain tumor dataset, they found that PBViT made great gains against standard CNN-based models. PBViT scores were 95.8\% accuracy, 95.3\% precision, 93.2\% recall, and 92\% F1-score which contained reasonable evidence to show its strength to find the tumors effectively. They built the model into a architecture that included DenseNet blocks and custom CNN layers together in the transformer to improve feature extraction and better represent tumor boundaries.

Simon and Briassouli\cite{b12} took an innovative approach to classification of brain tumors, using exclusively Transformer networks with a benchmarking study against CNNs. Their CNN model achieved an accuracy of 89.78\% while the ViT model achieved a higher accuracy of 96.5\%. Simon and Briassouli implemented dataset augmentation (to correct for the imbalance in the number of tumor categories) in order to improve the generalizability of the model. Asiri et al.\cite{b13} conducted a systematic study, using 5 finely-tuned pre-trained ViT models: R50-ViT-l16, ViT-l16, ViT-l32, ViT-b16, and ViT-b32. A robust assessment of performance metrics was used: precision, recall, f1-score, accuracy, and confusion matrices. Among the ViT models, the one with the best classification accuracy of 98.24\% was ViT-b32. The study recorded hyper-parameter tuning for all models: learning rate of 0.0001 and a batch size that aided convergence. The models were trained for 10 epochs using the Adam optimizer on a swell of medical images dataset.

Tariq et al.\cite{b14} proposed a deep learning-based multiclass classification method together with EfficientNetV2 and ViT architectures for detecting brain tumors. EfficientNetV2 model achieved 95\% accurate performance at a loss of 0.13. The F1-score, precision, and recall were all equal to 0.96 at this point. ViT model had a performance of only 90\% accuracy at a loss of 0.30. The F1-score, precision, and recall for the ViT model were all 0.89. The ensemble method proposed by the authors using geometrical mean was the best and achieved the most accurate prediction at 96\%. The dataset used in this study was retrieved from Kaggle and has been augmented using rotation, flipping, scaling, cropping, and adjustments for color. The inference was performed on an NVIDIA RTX 2080 GPU with an average timing of 0.35 seconds for each MRI image. When GPU memory was monitored, it was about 4.5 GB during inference and around 12 GB during training time frame. Overall the experimentation showed empirical feasibility for clinical use. Guo and Fan\cite{b15} applied a pre-trained ViT model for multi-label lung cancer classification on histological slices obtained from the LC25000 dataset in both a zero-shot and few-shot learning paradigm. The findings of the study showed strong performance under both approaches with the few-shot experiments achieving an accuracy of 99.87\% at epoch 1 and the ideal results reached at epoch 5.

Sun et al.\cite{b16} investigated the Swin Transformer model's use in lung cancer classification and segmentation. In terms of top-1 classification accuracy, the Swin-B model achieved 82.26\% exceeding ViT by 2.529\%. The image input pipeline segmented RGB lung CT image into non overlapping patches for linear embedding in C-dimensional space then run through the Swin Transformer blocks. Each Swin Transformer block consists of a shifted window-based multi-head self-attention (MSA) mechanism and multilayer perceptrons (MLPs) block both preceded by a layer normalization and succeeded by a residual connection. The Swin-B model was used for the majority of the work, while Swin-T and Swin-S models were only able to be explored in part due to not having the adequate hardware availability. Durgam et al.\cite{b17} proposed a framework termed Cancer Nexus Synergy (CanNS) for cancer diagnosis which primarily encompasses a Swin Transformer based unsupervised U-Net segmentation model (SwiNet), a Xception-LSTM GAN based on convolutional networks for classification (XLG), and a hyperparameter tuning method termed Devilish Levy Optimization (DevLO). The CanNS framework assessed using a Kaggle dataset is able to achieve a balance between the tradeoff of computational cost and accuracy for clinical application.
Chen and colleagues\cite{b18} designed an automated lung cancer detection framework employing a combination of CNN and Transformer architectures (specifically a Swin Transformer). The lung cell image segmentation was performed using Mask R-CNN prior to employing a Swin Transformer for classification. It was noted that this method of detection outperformed classical CNN based architectures such as ResNet50 to achieve a classification accuracy of 96.16\%. The experiments were conducted in Ubuntu 18.04.5 running PyTorch version 1.10.1 and using Python 3.6 and an NVIDIA GeForce RTX 2080Ti. The training schedule consisted of 100 epochs and a training - test split of 70\% - 30\% and a batch size of 12 and a learning rate decay at epochs 30, 60 and 80.

Akbari et al.\cite{b19} proposed a ViT-based method that focuses on the self-attention mechanism to extract essential features from histopathological images. The authors evaluated the overall classification results on two datasets LC25000 and IQ-OTH/NCCD, resulting in 98.80\%-99.09\% accuracy. The accuracy of the ViT model outperformed the performance metrics of CNN models based on comparison tests. The ViT model was trained using the Adam optimizer, with a learning rate of 0.000001 for 50 epochs, along with a batch size of 32 and used early stopping, model checkpointing, and learning rate scheduling. Extra data augmentation and preprocessing were accounted for using image data generators and Windows 10 operating system with Python-based frameworks using the Anaconda environment. Ouamane et al.\cite{b20} took a closer look at the various architectural parameters when using ViT models, including: i) patch size; ii) image resolution; iii) embedding dimension; iv) transformer depth; v) attention heads; and vi) size of MLP. The sub-model that obtained the highest accuracy, 99.77\%, on the PlantVillage dataset, used the following configurations: i) 224×224 image size; ii) patch size 16; iii) 512 embedding; iv) depth of 6; v) 8 heads; and vi) MLP size 1024.

Carolin and colleagues\cite{b21} investigated ViT architectures for skin lesion classification using pre-trained L16 and L32 models, in comparison with traditional classifiers such as decision trees, k-nearest neighbors, CNNs, and simpler ViT architectures. The ViT-L32 model reached an accuracy of 91.57\% with a melanoma recall of 58.54\% while the ViT-L16 returned 92.79\% accuracy with a melanoma recall of 56.10\%. Data augmentation was performed in the preprocessing of images, followed by training the model using transfer learning. Koushik and Karthik\cite{b22} used fine-tuned pre-trained models for COVID-19 detection using chest X-rays. The models included DenseNet, InceptionV3, WideResNet101, and ViT-B/32. The authors achieved a classification accuracy of 97.61\%, precision of 95.34\%, recall of 93.84\%, and F1-score of 94.58\%. All models were pre-trained using ImageNet, however, the original top layer of each model was removed and replaced by the output with regard to their dataset. They used the RectifiedAdam optimizer with ReduceLROnPlateau as their scheduler to enhance the performance of their models. Gelan and Se-woon\cite{b23} validated the success of their framework on two public ultrasound breast image analysis datasets: Mendeley and BUSI. Their framework returned AUC, MCC, and Kappa scores as a 1 ± 0 for the Mendeley dataset. They used BUViTNet to test the BUSI dataset, returning AUC of 0.968 ± 0.02, MCC of 0.961 ± 0.01, and Kappa score of 0.959 ± 0.02. BUViTNet outperformed ViT models trained from scratch, and ViT/CNN based transfer learning in breast ultrasound image classification.

\section{Background Study}
\subsection{Current research}
Breast cancer remains a critical global health challenge, having been accounted for the second leading cause of cancer-related deaths in the worldwide. Early detection is extremely important to help improve total survival by timely intervention, which really increases treatment outcomes. Over the years, imaging modality improvements have advanced from such modalities of mammography,
ultrasound, and thermography to the latest in diagnosis of breast cancer. Most importantly, mammography seems to be the gold standard in early detection together with the other imaging modality like MRI, which is the practice for high-risk individuals. Combined  with clinical examination or self-examination, these modalities are really quite important in picking the first appear of disease.

Recently, with AI, the deeply-rooted detection mechanism of breast cancer appears now rapidly changing. In various views, AI solutions are scanning images looking for abnormalities, different types of tumors and prognosis towards staging the cancers. Studies suggest detection rates of cancer in the range of 80\% to 90\% by mammography-based AI models\cite{b24}. By using mammography, it is quite possible to detect early breast cancer\cite{b25}. The algorithm will only identify abnormal masses to ease further investigation. The mammogram images used are taken from the mini MIAS database\cite{b26}. There is also improving interest in thermal infrared imaging suggesting non-invasive tests which have also been cleared from the nasty affliction of radiation upon the human body\cite{b27,b28,b29,b30}. This thermal imaging technique is known as thermography\cite{b31}. Here, the thermography clinic uses a thermal infrared camera. This camera clinic captures images picture of that region, which is considered\cite{b32}. Sometimes ultrasound, to a lesser extent than mammography, can load valuable texture information for segmenting solid nodules which leverage the so-called advanced algorithms such as 2D wavelet transforms\cite{b33}. In this paper, the authors applied the Pyramid Vision transformer(PVT v2) model, the first paper where the PVT v2 model is applied after realizing the model\cite{b34}. Here, the authors not only use PVT v2 but also four more vision transformer-based models and compare those models' results with this PVT v2 model, and PVT v2 achieves the highest accuracy than other models. The authors also compare three papers with this paper because they use the same dataset, but every paper's approaches are different. After comparing, the authors prove that their approaches are more advanced and their model's accuracy is better than those papers' results. A 3D hidden Markov model (HMM) is considered for breast tissue detection and tumor segmentation\cite{b35}.

\subsection{Our Approach}
Traditional deep learning models including CNNs and RNNs have become popular in breast cancer research. CNNs have been previously demonstrated to be highly successful in image-based tasks due to their ability to capture local patterns using convolutional layers. Since the models as such have fixed receptive fields, they are limited in regards to modeling long-range dependencies which are often essential for eight other complicated analyses of an image.

While transformers have been introduced for natural language processing, they increased their visibility as potential choices for image analysis. In this sense, the vision transformers provide the treatment of the image as a sequence of patches, much like words in a sentence. This architecture allows the transformer to attend to the global context and to model the intricate interdependencies within the data\cite{b36}. Compared to CNNs, transformers have better scalability and flexibility, justifying their selection as an alternative candidate for breast cancer diagnosis.

Taking inspiration from the successes of transformers across diverse domains, our research seeks to investigate their applicability toward breast cancer diagnosis based on mammographic images. Past works have relied heavily on CNN and other deep learning architectures, with far less examination into the innovative potential of the vision transformer. This very gap is being filled in this study; deploying vision transformers in clinical diagnosis under the challenging environment of data augmentation and loss optimization.

For comparative purposes in the evaluation of Swin Tiny, DeiT, BEiT, and ViT, along with YOLOv8, these have been trained on a dataset specific for cancerous and non-cancerous investigations. Assessment of different state-of-the-art architectures and performance measurement strategies will not only promote the advancement of breast cancer detection but also pave the way for further innovations in AI-based medical diagnostics.  

\section{Model Implementation and Dataset Augmentation Strategy}

Our study used a selected dataset from the Mendeley Data repository, encompassing a total of 745 original mammographic images and 9,685 augmented images\cite{b37}. The dataset had both cancerous and non-cancerous images well marked by expert clinicians to provide accuracy and reliability. Data augmentation techniques had been applied to overcome the limitations of the dataset and enhance the robustness of the model, which included random resizing and cropping to 224×224 pixels, horizontal flipping, and normalization using standard mean and standard deviation values.

\begin{figure*}[t]
\centering
\includegraphics[width=1\textwidth]{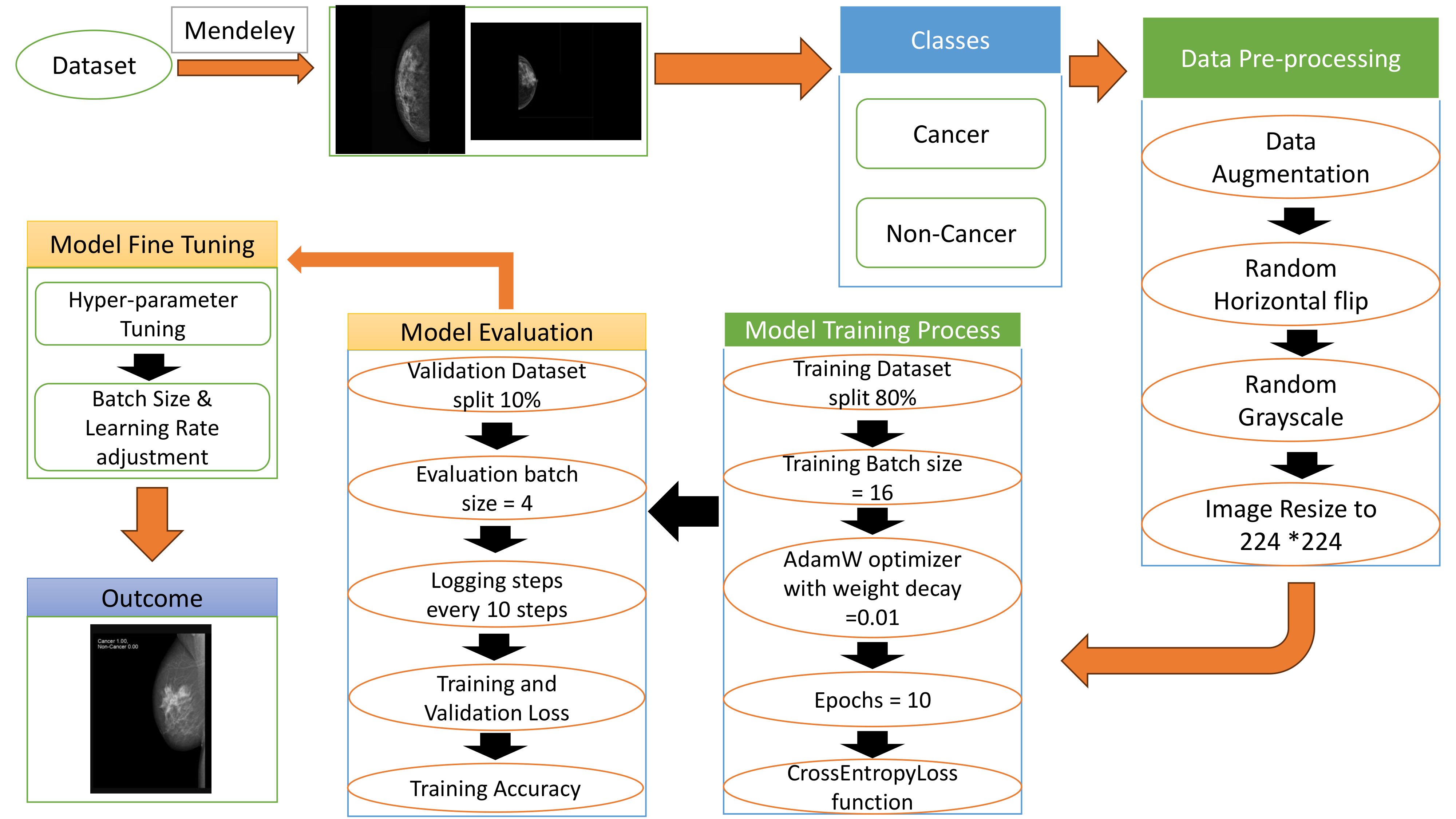}
\caption{Architecture of the proposed system design.}

\end{figure*}

We experimented with and evaluated five individual vision transformer models: Swin Tiny, DeiT, BEiT, ViT, and YOLOv8. Each of the models was set to have a batch size of 16 for training and a batch size of 4 for evaluation. The training was conducted for 10 epochs with a weight decay of 0.01. The evaluation strategy taken was epoch-based with logging every 10 steps to observe the model's performance.

The architecture of our proposed system in figure 1 leverages the global pattern recognition and complex dependencies among the image data. Our systematic training and intense evaluation sought to yield the most concessionary model in breast cancer detection, measuring performance using accuracy, training loss, and validation loss concerning the efficiency of each model. The setup allows the models to attain optimality, adjusted to various needs of the data set whilst setting the stage for future refinements in the diagnosis of breast cancer.

\section{Results and Analysis}
The effectiveness of five vision transformer models, Swin Tiny, DeiT, BEiT, ViT, and YOLOv8, in observing breast cancer in mammographic images was tested with the organized dataset. Table 1 summarises the training and validation loss, as well as precision attained for each model, thus imparting information about them.

\begin{table}[ht]
\centering
\caption{Overall Model Performance Results}
\label{tab:1}
\begin{tabularx}{\columnwidth}{|*{11}{X|}} 
\hline
Model & Training Loss & Validation Loss & Accuracy  \\ \hline
Swin Tiny & 0.179700 & 0.049648 & 0.979866  \\ \hline
DeiT & 0.160800 & 0.042177 & 0.986577 \\ \hline
BEiT & 0.285400 & 0.060966 & 0.979866 \\ \hline
ViT & 0.228000 & 0.086377 & 0.993289 \\ \hline
YOLOv8 & 0.3907 & 0.489 & 0.97 \\ \hline
\end{tabularx}
\end{table}

ViT outperformed the rest with the maximum accuracy of 99.32\%, but with slightly higher validation loss (0.0863) than DeiT whereas its performance outlines a strong generalization power over the dataset, thus appearing a favorable candidate for breast cancer detection. Close on its heels was DeiT with 98.66\% accuracy, together with the least training loss (0.1608), and validation loss (0.0421), indicating itself as efficient in data learning. Likewise, both Swin Tiny and BEiT delivered mediocre results, whereas the two scored around 97.9\% but BEiT gained much higher loss values, some optimization is in demand.

On the contrary, although reaching an impressive accuracy of 97\%, YOLOv8 was subjected to the highest training loss (0.3907) and validation loss (0.489), conveying clear inefficiencies in working with the dataset. This reflects its limitations in tasks needing fine-grained feature extraction and robust learning capabilities.

\begin{figure}[htbp]
  \centering
  \begin{minipage}[b]{1\linewidth}
    \centering
    \includegraphics[width=1\linewidth]{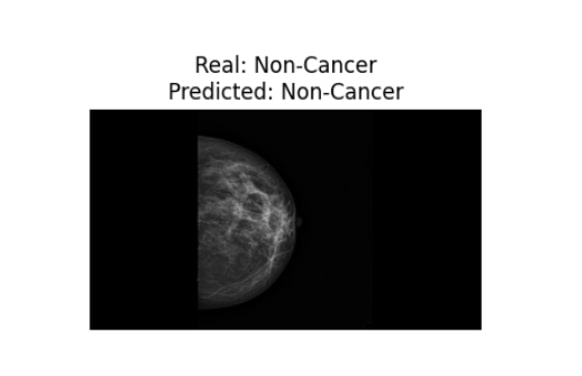}
  \end{minipage}
  \hfill
  \begin{minipage}[b]{1\linewidth}
    \centering
    \includegraphics[width=1\linewidth]{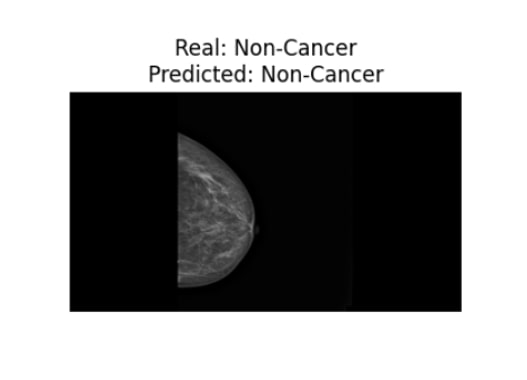}
  \end{minipage}
  \caption{After analyzing images from both cancerous and non-cancerous cases}

\end{figure}

In Figure 2, we can show the images that we get after training our models which is able to detect Cancer and Non-cancer images.

\begin{figure}[htbp]
  \centering
  \includegraphics[width=1\columnwidth]{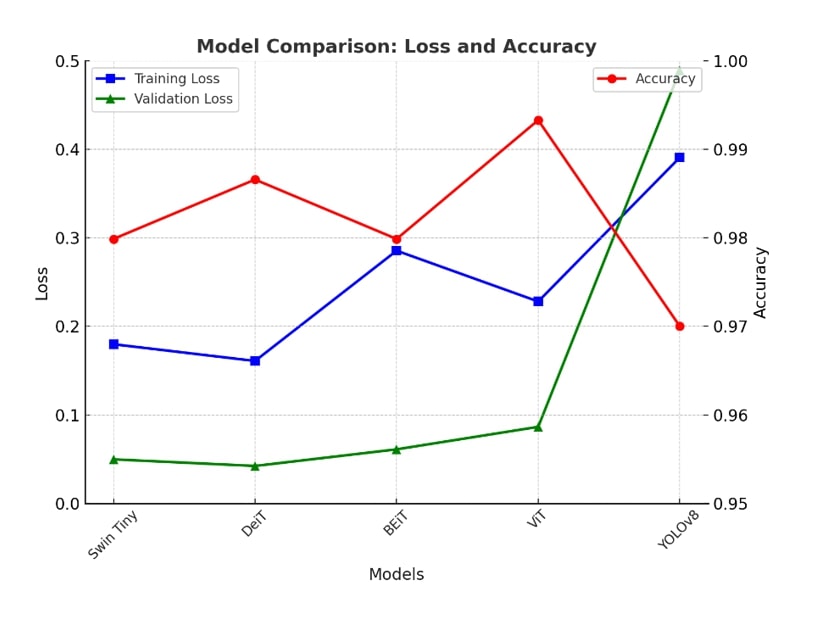} 
  \caption{Comparison of Training and Validation Loss with Accuracy Across Different Models.}
  \label{fig:j}
\end{figure}

Training and validation loss graphs for all models, demonstrating an overall trend of error reduction across epochs, are depicted in Figures 3, thus confirming that the two best-performing models were ViT and DeiT; ViT yields marginally higher overall accuracy than DeiT. This is a clear testimony of the deep learning aspect, suggesting that the vision transformer is a strong candidate for breast cancer detection, especially for tasks that are dependent upon delicate visual features and complex dependencies. When studies in the literature are examined, besides the accuracy measurementmetric, precision (Prec), sensitivity (Recall), and F1 score (F1) metrics were observed to be used. These values can be calculated in matrix form using the confusion matrix. The true positive values (TP), true negative (TN), false positive (FP) and false negative (FN) can be calculated in the confusion matrices of the classification results. Table II shows the components of the confusion matrix 

\begin{table}[h]
\centering
\caption{Confusion matrix.}
\begin{tabular}{c|c|c|c}
\multirow{2}{*}{\textbf{Actual Value}} & \multicolumn{2}{c|}{\textbf{Predicted Value}} & \\
\cline{2-3}
 & \textbf{Positive} & \textbf{Negative} & \\
\hline
\textbf{Positive} & True Positive (TP) & False Negative (FN) & \\
\hline
\textbf{Negative} & False Positive (FP) & True Negative (TN) & \\ \hline
\end{tabular}

\vspace{1em}
\[
\text{Accuracy} = \frac{TP + TN}{TP + FP + FN + TN}
\]
\[
\text{Precision} = \frac{TP}{TP + FP}
\]
\[
\text{Recall} = \frac{TP}{TP + FN}
\]

\[
\text{F1--Score} = \frac{2 \times \text{Precision} \times \text{Recall}}{\text{Precision} + \text{Recall}}
\]
\end{table}

A confusion matrix is a table that is often used to numerically determine the perfor-mance of a classification method on a test dataset where the actual values are known. The results of the confusion matrix on the data sets we used in the study are as follows: 

\begin{itemize}
    
\item TP (True Positive): It means cancer.
\item FP (False Positive): It means non-cancer cases as cancer.
\item TN (True Negative): It means non-cancer.
\item FN (False Negative): It means cancer case as non-cance.
  
\end{itemize}

\begin{figure}[htbp]
  \centering
  \includegraphics[width=1\columnwidth]{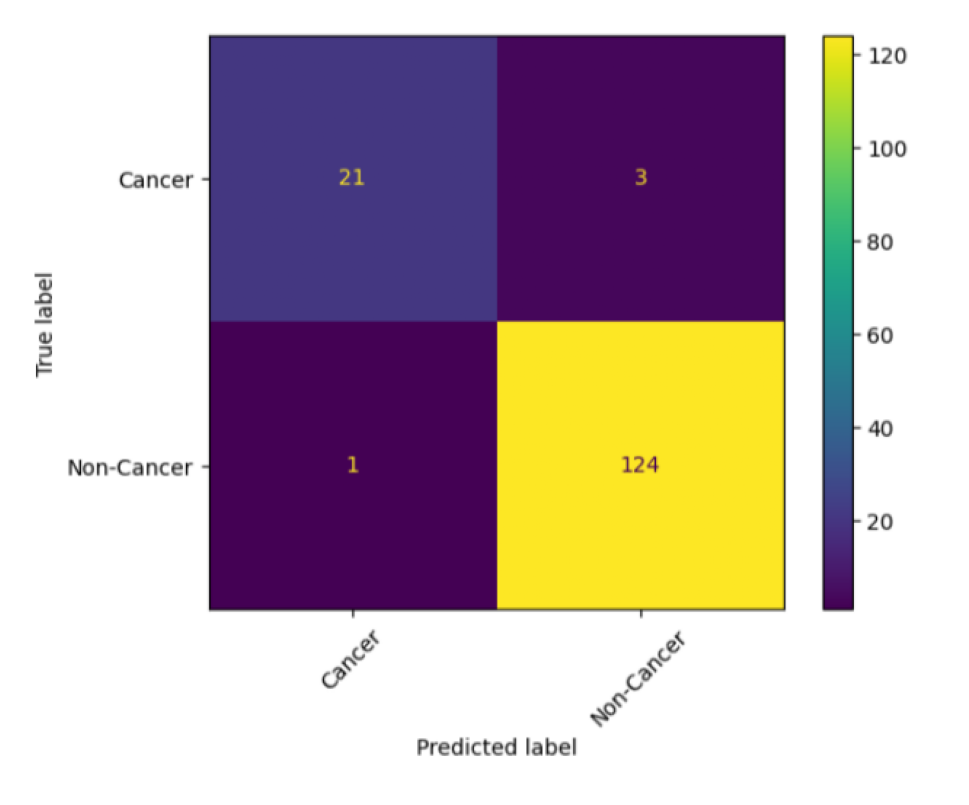} 
  \caption{Confusion Matrix for Breast Cancer Detection Using ViT.}

\end{figure}

Figure 4 tritely the performance of the ViT model for classification of breast cancer cases into "Cancer" and "Non-Cancer" categories. The confusion matrix accents the fact that the model rightly classified 21 "cancer" cases (True Positives) and 124 "non-cancer" cases (True Negatives), on one hand, but wrongly classified three "non-cancer" cases as cancer (False Positives) and one "cancer" case as non-cancer (False Negative) on the other. It is thus true that the model's overall performance is quite strong, and thus the key metrics calculated from the matrix show that the model is very effective.

Results lend support to the point that the transformer-based method could display high diagnostic accuracy. The higher loss values on some of the models like YOLOv8 and BEiT suggest that there is certainly a chance of improvement through better optimization, such as better hyperparameter tuning and augmentation techniques. These findings lay some solid groundwork toward future work that intends to make AI-driven diagnostic techniques much better. 

\section{Comparative Analysis with Previous Work Using the Breast Cancer Dataset}

Liang et al.\cite{b38} described a novel convolutional neural network (CNN) architecture for breast cancer classification in which both 2D and 3D mammograms are incorporated concurrently. The experimental results showed that using two forms of imaging leads to higher classification performance. Liang et al.'s model utilizes three CNN classifiers, resulting in a model with an AUC score of 0.97, which is an increase of 34.72\% when compared to classifiers that used a single imaging modality. The proposed network was comprised of two networks: a backbone CNN for feature extraction and a shallow CNN classifier that used the features from the previous network. For the study data, digital mammography (DM) and digital breast tomosynthesis (DBT) data was retrospectively collected from patients at the University of Kentucky Medical Center from January 2014 to December 2017. The study used the 2D-T3-Alex and 3D-T2-Alex as the baseline methods for DM and DBT analyses respectively. The 2D-T3-Alex model uses a pre-trained AlexNet for feature extraction, while the 3D-T2-Alex model uses AlexNet to extract feature maps for each DBT slice and inputs these feature maps into a single layer CNN for classification where K=30 feature maps were taken per DBT.

Seyfullah and colleagues\cite{b39} proposed a new method of breast cancer detection by using thermal images through the use of four well-adopted deep neural networks that have shown effectiveness on object classification. Of the four, ResNet50 had the best classification performance for breast cancer detection with a test accuracy of 88.89\%. Prinzi et al.\cite{b40} tested three different YOLO models, YOLOv3, YOLOv5 and YOLOv5-Transformer for detecting breast cancer. They also included Eigen- CAM for model explainability, to highlight suspicious areas on mammograms. The small YOLOv5 model had the highest performance with an mAP (mean Average Precision) detecting breast cancer on their custom dataset with 0.621 mAP. They also used mosaic data augmentation to help with detection of multiple lesions on a single image and created a 2 × 2 image tile so that the target image would be combined with three random images in the tile. Sadia et al.\cite{b41} used machine learning methods Support Vector Machine (SVM), Logistic Regression (LR), and K-Nearest Neighbor (KNN) for breast cancer classification. The detection model took 97.7\% accuracy with an FPR of 0.01, FNR of 0.03, and an AUC of 0.99 with a training fraction of 60\% and a test fraction of 40\%. This study shows that minimizing Type I and Type II error could lead to improved sensitivity and accuracy.

Manav and his team\cite{b42} compared different machine learning algorithms including SVM, Decision Tree, Naive Bayes (NB), KNN, AdaBoost, XGBoost and Random Forest using the Wisconsin Breast Cancer Dataset. Their purpose was to evaluate classification performance based on the four classification metrics: accuracy, precision, sensitivity, and specificity. Upon testing with XGBoost their findings showed the algorithm produced the highest accuracy of 98.24\% while also having the lowest error rate. Tahmooresi et al.\cite{b43} proposed a hybrid model of several major machine learning algorithms, SVM, Artificial Neural Network (ANN), KNN, and Decision Tree, for accurate detection of breast cancer. The group found SVM achieved the most successful accuracy at 99.8\% for breast cancer detection, although researchers suggested that performance could be increased to 100\%. Majid et al.\cite{b44} even went further where it was possible to classify breast tumors into subcategories, such as Fibroadenoma, or specifically Lobular Carcinoma, instead of only benign or malignant tumors. The DenseNet Convolutional Neural Network (CNN) that produced a multiclass object detection method using the BreakHis histopathology image dataset achieved 95.4\% accuracy in classification. The final average image-level accuracy produced from their model was 95.4\% and 96.48\% for patient-level classification across all magnification factors.

Mümine Kaya Keleş\cite{b45} stated she wanted to make it possible for early detection of breast cancer using non-invasive and painless measures through data mining algorithms. A comparative study was made on classification algorithms that exist in Weka on a dataset consisting of antennas that provided measures for frequency bandwidth, dielectric constant, electric field, and tumor data. The algorithms Bagging with IBk, Random Committee, Random Forest, SimpleCART each gave measures of accuracy that were over 90\% when validated with a 10fold cross-validation. Asaf, and others, proposed a new deep learning framework, DeepBreastCancerNet, consisting of 24 layers, with six convolutional layers, nine inception modules, and one fully connected layer. Clipped ReLU, leaky ReLU, batch normalization, and cross-channel normalization were used in their architecture for DeepBreastCancerNet\cite{b46}. This model had a classification accuracy of 99.35\%, and it performed better than other existing deep learning models, with 99.63\% validation accuracy on a publicly available dataset. Wolfgang, and others, used cancer detection in a study using subjects that included 867 patients with palpable or mammographically detected index lesions\cite{b47}. The cancer detection prevalence detected by screening sonography was 0.41\% which comprised 22\% of all nonpalpable cancers. The mean size of invasive cancers only detected by sonography was 9.1 mm which was on par with cancer sizes identified through mammography. The classification from sonography yielded 100\% sensitivity and 31\% specificity.

Bibhuprasad et al.\cite{b48} presented a hybrid model with artificial intelligence methods, combined with extensions of multivariate statistical methods. The model included Principal Component Analysis (PCA) for selection of dimensions or features in variables and also employed Artificial Neural Networks (ANN) for classifying the data. The Ohio State University computer laboratory used the Wisconsin Breast Cancer dataset to evaluate the model with a 10-fold cross-validation and classification was carried out against different classifiers. Tanzila et al.\cite{b49} described a method of shape based feature extraction in order to classify tumor cells as benign or malignant. This method used Naive Bayes and ANN (Artificial Neural Network) classifiers and exhibited a grading stage for malignant cells. Cross validation for this was conducted with a classification recall of 98\%, which is better than conventional physical measurements. Finally, Yousuf et al.\cite{b50} exhibited a three-way process for diagnosing breast tumors. The first stage involved Fuzzy c-Means clustering

\clearpage 
\onecolumn
\renewcommand{\arraystretch}{1.15}
\begin{longtable}{|p{3cm}|p{1.2cm}|p{4cm}|p{2cm}|p{5cm}|}
\caption{ANALYZING PERFORMANCE MEASURES IN COMPARISON TO OTHER TECHNIQUES} \\
\hline
\textbf{Study} & \textbf{Year} & \textbf{Methodology} & \textbf{Accuracy} & \textbf{Notable Contributions} \\
\hline
\endfirsthead

\multicolumn{5}{c}%
{{\bfseries \tablename\ \thetable{} -- continued from previous page}} \\
\hline
\textbf{Study} & \textbf{Year} & \textbf{Methodology} & \textbf{Accuracy} & \textbf{Notable Contributions} \\
\hline
\endhead

\hline \multicolumn{5}{|r|}{\textit{Continued on next page}} \\
\hline
\endfoot

\hline
\endlastfoot

\hline

Gongbo et al. & 2020 & CNN-based fusion model using three CNN classifiers for 2D/3D modalities & 97\%  & First to jointly leverage 2D and 3D mammogram data in a unified CNN framework\\ \hline
Seyfullah et al. & 2019 & Four deep CNNs, including ResNet‑50 & 88.89\% & Novel application of deep CNNs to thermal imaging for breast cancer, with architecture benchmarking \\ \hline
Prinzi F. et al. & 2023 & YOLOv3, YOLOv5, YOLOv5- 
Transformer; transfer learning on private data; Eigen‑CAM interpretability & mAP = 0.621 (small YOLOv5) & YOLO comparison in mammography \\ \hline
Sadia et al. & 2022 & SVM, KNN, Logistic Regression & 97.7\% & Low error rates, improved preprocessing, model comparison, clinical focus \\ \hline
Manav et al. & 2022 & Employed seven machine learning algorithms & 98.24\% & Its comprehensive comparison of multiple machine learning algorithms on the same dataset, providing a clear understanding of each model's strengths and weaknesses in the context of breast cancer detection.\\ \hline
M. Tahmooresi et al. & 2018 & The study proposes a hybrid model combining several Machine Learning algorithms such as Support Vector Machine, Artificial Neural Network, K-Nearest Neighbor, and Decision Tree & Does not specify the exact accuracy & The integration of multiple ML algorithms into a hybrid model capable of handling diverse data types \\ \hline
Majid et al. & 2018 & Utilized a Convolutional Neural Network (CNN) model, specifically DenseNet architecture with three dense blocks and transition layers. & 95.4\% & Extends beyond binary classification to classify specific subtypes of breast tumors, such as Fibroadenoma and Lobular carcinoma, using deep learning techniques.\\ \hline
Mümine Kaya Keleş & 2019 & A 10-fold cross-validation method was used to evaluate the performance of the algorithms. & 90\% & Its comparative analysis of multiple data mining classification algorithms using a unique dataset derived from antenna measurements for breast cancer detection\\ \hline
Asaf et al.  & 2023 & Proposed a custom deep learning model named DeepBreastCancerNet comprising 24 layers: 6 convolutional layers, 9 inception modules, and 1 fully connected layer. & 99.63\% & The model's design, including inception modules and specific activation functions, enhances feature extraction and classification performance.\\ \hline
Wolfgang et al. & 2000 & High resolution sonography performed as an adjunct to mammography. Lesion Classification: Sonographically detected lesions were prospectively classified into benign, indeterminate, or malignant categories & Sensitivity: 100\% for detecting malignancy. Specificity: 31\%. Cancer Detection Rate: 0.41\% overall; sonography detected 22\% of nonpalpable cancers & Demonstrated that high-resolution sonography can detect clinically and mammographically occult breast cancers, particularly in women with dense breast tissue.\\ \hline
Bibhuprasad et al. & 2019 & Proposed a hybrid model combining Principal Component Analysis (PCA) for feature selection and Artificial Neural Network (ANN) for classification.  & 97\%  & Integrating PCA for effective feature selection with ANN for classification, aiming to improve diagnostic accuracy in breast cancer detection. \\ \hline
Tanzila et al. & 2019 & Proposed a cloud-based decision support system that utilizes shape-based features extracted from breast cytology images for the detection of tumor cells. & 98\% & This work lies in the integration of a cloud-based framework with shape-based feature extraction and classification techniques (Naive Bayesian and ANN) for the detection and classification of malignant cells in breast cytology images\\ \hline
Yousif et al. & 2018 & Methodology involves several steps: image enhancement using adaptive median filtering and Balance Contrast Enhancement Technique (BCET), segmentation using thresholding and Fuzzy C-Means (FCM), feature extraction using Discrete Wavelet Transform (DWT), and classification using a Probabilistic Neural Network (PNN).  & 90\% & Integration of multiple image processing techniques (adaptive median filtering, BCET, FCM, DWT) with a Probabilistic Neural Network for accurate detection and classification of breast cancer from mammography images. \\ \hline
Basem et al. & 2023 & Proposed a deep learning model named BCCNN for the detection and classification of breast cancer into eight classes. Employing five pre-trained deep learning models (Xception, InceptionV3, VGG16, MobileNet, and ResNet50) fine-tuned on the dataset & 98.287\%  & Integration of data augmentation using GANs with a custom-designed deep learning model (BCCNN) and the comparative analysis with five fine-tuned pre-trained models. \\ \hline
Asmaa A. et al. & 2021 & Automated optimal Otsu thresholding to identify tumor-like regions (TLRs). Feature Extraction: Deep Convolutional Neural Networks (CNNs) using AlexNet and ResNet-50 architectures. Classification: Support Vector Machine (SVM) to classify mammogram structures into four categories. &  91\% & Integration of deep CNNs with SVM for multi-class classification of mammogram images. Application of automated optimal Otsu thresholding for TLR identification. \\ \hline

\end{longtable}
\clearpage 
\twocolumn
\vspace{10pt} 

with thresholding in form of image segmentation. The second stage used the Discrete Wavelet Transform (DWT) for feature extraction, while the final stage relied on Artificial Neural Networks with Benign, Malignant, and Normal output classification. Additionally, automated classification utilized Probabilistic Neural Networks (PNN) using radial basis functions for additional data, to allow for early detection overall.

Basem et al.\cite{b51} ran 30 experiments using a custom deep learning model and five pre-trained models working on several datasets. The performance metrics calculated were F1-score, recall, precision, and accuracy. The proposed Breast Cancer CNN (BCCNN) achieved the highest F1-score of 98.28\% for the model class proposed when compared to the other models analysed, which were ResNet50 (98.14\%), Xception (97.54\%), and others. Asmaa et al.\cite{b52} introduced a computer-assisted diagnosis system that uses deep CNNs called AlexNet and ResNet-50 to process extracted TLRs from mammograms. The CNN features were normalised and then classified by a Support Vector Machine (SVM). Under five-fold cross-validation the system reported achieving an accuracy of 0.91 with AlexNet and 0.84 with ResNet-50 on four class ROIs. 

However, in our research, we took a dataset from Kaggle applied vision transformer-based models and compared those models' training loss with the Yolo v8 model. For fine-tuning, we made the same parameter changes for all models. Still, as a result, we noticed that the DeiT model training loss is lower than that of other models, but the accuracy of the ViT model is higher than that of different models. Regarding training loss, validation loss, and accuracy, the Yolo v8 model performance is lower than other vision transformer-based models.

\section{Limitation and Future Work}
Our study has some limitations, such as dependence on a single dataset which may not capture the diversity of the population, and evidence of significant training and validation loss thereby calling for further optimization. Inasmuch as, the other crux for the deployment of transformer models are with respect to enormous computational resources they require, their application in resource-constrained settings remains limited. Future work should augment datasets with diverse populations and multi-modal data such as MRI and patient histories. Lightweight models should be built for possible deployment into low-power computing devices and should be validated clinically with the oncology profession. Ethical issues related to data confidentiality and bias must be addressed if AI-led breast cancer detection approaches are to elicit trust and adaptability.

\section{Conclusion}
Breast cancer is one of the significant causes of death among women, with one woman affected by breast cancer out of eight women\cite{b5}. In the diagnosis process, due to the wide range of features associated with breast abnormalities, some abnormalities may be missed or misinterpreted. Breast cancer diagnosis can be highly accurate; it is not necessarily the same as the results obtained in other sets of images. Future research can be done by improving the system’s performance and validating it by conducting tests on a more extensive set of images. Vision transformers have emerged as robust architectures for learning long-range dependencies over the last two years. However, their potential in multi-view mammogram analysis remains unleashed mainly to date. In this study, we applied different types of vision transfer learning models, which can be a novel approach to detecting breast cancer. Here, we presented an automatic segmentation method based on machine learning algorithms for breast ultrasound images. The automatic segmentation results demonstrate good consistency with visually manually segmented ground truth. We build this AI so that women get proper treatment to prevent breast cancer, and in this way, they can improve their life quality.

\vspace{12pt}
\color{red}

\end{document}